\documentstyle[12pt]{article}
\begin{document}
\pagestyle{empty}
\begin{flushright}
   {CERN-TH.6299/91} \\
\end{flushright}
\vspace* {10mm}
\renewcommand{\thefootnote}{\fnsymbol{footnote}}
\begin{center}
   {\bf A TWO PARAMETER DEFORMATION OF THE $SU(1/1)$ SUPERALGEBRA
   AND THE XY QUANTUM CHAIN IN A MAGNETIC FIELD} \\[13MM]
   Haye Hinrichsen \\[5mm]
   {\it Universit\"{a}t Bonn,
   Physikalisches Institut \\ Nussallee 12, W-5300 Bonn 1, FRG} \\[10mm]
   Vladimir Rittenberg\footnote{Permanent address:
   Universit\"{a}t Bonn,
   Physikalisches Institut, Nussallee 12, W-5300 Bonn 1, FRG} \\[5mm]
   {\it Theory Division, CERN \\
   CH-1211 Geneva 23, Switzerland} \\[4cm]
	  {\bf Abstract}
\end{center}
\renewcommand{\thefootnote}{\arabic{footnote}}
\addtocounter{footnote}{-1}
\vspace*{2mm}
We show that the XY quantum chain in a magnetic field is invariant under a two
parameter deformation of the $SU(1/1)$ superalgebra. One is led to an extension
of the braid group and the Hecke algebra which reduce to the known ones when
the two parameter coincide. The physical significance of the two parameters is
discussed.
\vspace{3cm}
\begin{flushleft}
   CERN-TH.6299/91 \\
   October 1991 \\
\end{flushleft}
\thispagestyle{empty}
\mbox{}

\newpage
\setcounter{page}{1}
\pagestyle{plain}

There were several attempts to extend the one-parameter quantum algebras to
multiparameter ones \cite{r1}. As shown however by Reshetikhin \cite{r2} the
link polynomials depend only on one parameter. One can state this result in a
different way: if one has a one-dimensional quantum chain which is invariant
under a multiparameter quantum algebra, one can do a similarity transformation
which eliminates all the parameters but one. As will be shown in this paper,
the situation is different in the case of quantum superalgebras. We will start
with a physical example. Consider the quantum chain

\begin{equation}
H\;=\;\Omega_{q}\sum_{i=1}^{L}\sigma_{i}^{z}\;
+\;\frac{\Omega_{\eta}}{2}\sum_{i=1}^{L-1}\:[(1+u)\sigma_{i}^{x}\sigma_{i+1}^{x}
+\: (1-u)\sigma_{i}^{y}\sigma_{i+1}^{y}]\;+\;B\;+\;S,
\label {e1}
\end{equation}

\noindent
where $\sigma^{x}$, $\sigma^{y}$ and $\sigma^{z}$ are Pauli matrices
inserted in the i-th position of the Kronecker product

\begin{eqnarray}
& & \sigma^{k}_{i}\;=\;\bf{1}\otimes\bf{1}\otimes\ldots\otimes
\underbrace{\sigma^{k}}_{i} \otimes
\ldots\otimes\bf{1}\otimes\bf{1} \;\;\;\;\;\;\;\;\;\;\;\;\;\;
\rm{(i=1,2,\ldots L)}
\label{e2}
\\ \,
& & [\sigma_{i}^{k},\sigma_{j}^{l}]\; = \;0.
\;\;\;\;\;\;\;\;\;\;\;\;\;\;\;\;\;\;\;\;\;\;\;\;\;\;\;\;\;\;\;\;\;\;\;\;\;\;\;
\;\;\;\;\;\;\;\;\;\;\;\;\;\;
 \rm{(i\neq j)}\nonumber
\end{eqnarray}

\noindent
$\Omega_{q}$, $\Omega_{\eta}$ and $u$ are parameters, $B$ and $S$ are boundary
and surface terms respectively. This chain appears in the domain wall theory
of two-dimensional commensurate-incommensurate phase transitions \cite{r3}
and in Glauber's kinetic Ising model \cite{r4}. We will first make an important
change of notations, choose $B=0$ (no periodic boundary conditions!)
and fix $S$ by

\begin{eqnarray}
\Omega_{q}\;=\;\frac{q+q^{-1}}{2},\;\;\;\;\;\;
& & \Omega_{\eta}\;=\;\frac{\eta+\eta^{-1}}{2},\;\;\;\;\;\;\;\;\;\;\;
u\;=\;\frac{\eta-\eta^{-1}}{\eta+\eta^{-1}}
\label{e3} \\ \, \nonumber \\
& S & \;=\;\frac{1}{2}\:(-q^{-1}\sigma_{1}^{z}-q\:\sigma_{L}^{z}).\nonumber
\end{eqnarray}

\noindent
With this change of notations we have

\begin{eqnarray}
& & H \; = \; H(q,\eta) \; = \; \sum_{i=1}^{L-1}\:H_{i}(q,\eta)
\label {e4}
\\
& & H_{i}(q,\eta) \; = \; \frac{1}{2}\: [\eta\:\sigma_{i}^{x}\sigma_{i+1}^{x}
\; + \; \eta^{-1}\sigma_{i}^{y}\sigma_{i+1}^{y}
\; + \; q\,\sigma_{i}^{z} \; + \; q^{-1}\sigma_{i+1}^{z} ].\nonumber
\end{eqnarray}

A detailed discussion of the properties of the chain given by eq. (\ref{e4})
will be given elsewhere~\cite{r5}, here we are going to mention only a few.
First, there are the symmetry properties

\begin{equation}
\label{e5}
H(q,\eta) \; \doteq \; H(q^{-1},\eta) \; \doteq \; H(q,\eta^{-1})
\; \doteq \; H(\eta,q).
\end{equation}

The ''equality'' among the Hamiltonians implies that the spectra are identical.
The first two equalities are obvious but not the last one which reminds of
duality transformations of quantum chains \cite{r6}. In the continuum limit,
one has the following phase structure \cite{r3,r5}:

\begin{tabbing}
\ \ \ \ \ \ \ \ \ \= $\Omega_{q}\leq 1$,\ \ \  $\Omega_{\eta}\leq 1$:\ \ \
\ \ \ \ \ \ \ \ \ \ \ \ \ \ \ \ \ \ \ \ \ \ \  \ \ \ \ \
\=massless-incommensurate
\\

\>$\Omega_{q}\leq 1$,\ \ \  $\Omega_{\eta}>1$\ or \ $\Omega_{q}>1$,\ \ \
$\Omega_{\eta}\leq 1$:\ \ \ \> massive incommensurate \\

\>$\Omega_{q}>1$,\ \ \  $\Omega_{\eta}>1$,\ \ \ $\Omega_{q} \neq
\Omega_{\eta}$:
\ \ \> massive\\

\>$\Omega_{q}=\Omega_{\eta}$, \ \ \ $(\Omega_{q} > 1)$ : \
\ \ \>critical Ising type \\

\>$\Omega_{q}=\Omega_{\eta}=1$ : \ \ \>Pokrovsky-Talapov phase transition \\
\end{tabbing}

It is by now clear that the properties of the chain depend on \underline{both}
parameters $q$ and $\eta$. \\ We now perform a Jordan-Wigner transformation.
First write $\sigma_{j}^{z}\;=\;-i\sigma_{j}^{x}\sigma_{j}^{y}$ and next define

\begin{eqnarray}
\label{e6}
& & \tau_{j}^{x,y}\;=\;\exp({\frac{i \pi}{2} \sum_{k=1}^{j-1}
(\sigma_{k}^{z}+1)})
\;\sigma_{j}^{x,y} \\
& & \{\tau_{i}^{x},\tau_{j}^{x}\} \;=\;
    \{\tau_{i}^{y},\tau_{j}^{y}\} \;=\;
    \{\tau_{i}^{x},\tau_{j}^{y}\} \;=\; 0.
\ \ \ \ \ \ \ \ \ \ \ \ \ \ \ \ \ \ (i \neq j) \nonumber
\end{eqnarray}

\noindent
Using Eq. (\ref{e4}) and (\ref{e6}) we get

\begin{eqnarray}
& & H_{j} \;=\; \frac{1}{2} [\:\eta\tau_{j}^{x}\tau_{j+1}^{x}\: +
\:\eta^{-1}\tau_{j}^{y}\tau_{j+1}^{y} \: -
\: i q \tau_{j}^{x}\tau_{j}^{y} \: - \: i q^{-1}\tau_{j+1}^{x}\tau_{j+1}^{y}].
\end{eqnarray}

\noindent
We now observe the following important identity

\begin{equation}
\label{e7}
[T^{X},H(q,\eta)] \;=\; [T^{Y},H(q,\eta)] \;=\; 0
\end{equation}

\noindent
with

\begin{eqnarray}
\label{v9}
& & T^{X} \;=\; \Delta(\tau^{x}) \;=\; \sum_{j=1}^{L}\:
    \alpha^{\frac{1-L}{2}} \;\sum_{j=1}^{L} \alpha^{j-1} \tau_{j}^{x}
\nonumber \\
& & T^{Y} \;=\; \Delta(\tau^{y})\;=\;\sum_{j=1}^{L}\:
    \beta^{\frac{1-L}{2}} \;\sum_{j=1}^{L}  \beta^{j-1} \tau_{j}^{y}
\\
& & \nonumber
\\
\label{v10}
& & \{T^{X},T^{X}\} \;=\; 2[L]_{\alpha}, \ \ \ \ \ \
    \{T^{Y},T^{Y}\} \;=\; 2[L]_{\beta}, \ \ \ \ \ \
    \{T^{X},T^{Y}\} \;=\; 0,
\end{eqnarray}

\noindent
where $L$ is the length of the chain and

\begin{equation}
\label{e9}
\alpha \;=\; -\frac{q}{\eta}, \ \ \ \ \ \ \ \
\beta  \;=\; -q \eta, \ \ \ \ \ \ \ \
[L]_{\lambda} \;=\; \frac{\lambda^{L}-\lambda^{-L}}{\lambda-\lambda^{-1}}.
\end{equation}

The equalities (\ref{e7}) come from the existence of a fermionic zero mode for
a
$q$ and $\eta$. The equations (\ref{v10}) togehter with the coproduct
(\ref{v9})
give a representation of a Hopf algebra. Before we proove this statement let us
consider the case $\alpha=\beta=-q$. We first notice that in this case
$S^{z}=\frac{1}{2}\:\sum_{i=1}^{L}\sigma_{i}^{z}$ also commutes with
$H(q,\eta)$
We now remind the reader the $U_{\alpha}[SU(1/1)]$ algebra \cite{r7}.
With $A^{\pm}= \frac{1}{2}(T^{X}\pm iT^{Y})$ we have

\begin{eqnarray}
\label{e10}
& & \{A^{\pm},A^{\pm}\}=0, \ \ \ \ \ \ \{A^{+},A^{-}\}=[E]_{\alpha}, \ \ \ \ \
\ \,[S^{z},A^{\pm}] = \pm A^{\pm} \\ & & [E,S^{z}]=[E,A^{\pm}]=0 \nonumber
\end{eqnarray}

\noindent
with the coproduct

\begin{eqnarray}
\label{e11}
& & \Delta(\alpha,A^{\pm}) \;=\; \alpha^{E/2} \otimes A^{\pm} +
       A^{\pm}\otimes\alpha^{-E/2}  \nonumber \\
& & \Delta(\alpha,S^{z}) \;\;=\; S^{z}\otimes {\bf1} + {\bf1}\otimes S^{z} \\
& & \Delta(\alpha,E) \;\;\;=\; E \otimes {\bf1} \; + {\bf1} \otimes E.
\nonumber
\end{eqnarray}

The fermionic representations correspond to take $E={\bf1}$, \
$S^{z}=\frac{1}{2}\sigma^{z}$, \ $A^{\pm}=a^{\pm}$ and $\{a^{+},a^{-}\}=1$ in
Eq
(\ref{e11}). In this representation $E$ in Eq. (\ref{e10}) is equal to $L$ (the
number of~sites). Comparing now (\ref{e10}), (\ref{e11}) with
Eqs. (\ref{v9},\ref{v10}) we
observe \cite{r8} that the quantum chain (\ref{e4}) with $\eta=1$ is invariant
under $U_{\alpha}[SU(1/1)]$ transformations. It was also shown by Saleur
\cite{r8}
that the quantities $U_{j}=\Delta_{q}-H_{j}(q,1)$ are the generators of the
Heck
algebra
\begin{eqnarray}
\label{e12}
& & U_{j}^{2} \;=\; 2\:\Delta_{q}\:U_{j} \nonumber \\
& & U_{j}U_{j\pm 1}U_{j} - U_{j} \;=\; U_{j\pm 1}U_{j}U_{j \pm 1} - U_{j \pm 1}
\\
& & U_{i}U_{i\pm j} \;=\; U_{i \pm j}U_{i}. \ \ \ \ \ \ \ \ \ \ \ \ \ \
\ \ \ \ \ \ \ \ \ \ \ \ \ \ \ \ \ \ \ \ \ \ \ \ \ \ \ \ \ \ \ (j \neq 1)
\nonumber
\end{eqnarray}

\noindent
Actually they correspond to a quotient of this algebra since the generators
satisfy also the relations \cite{r9}

\begin{equation}
\label{e13}
U_{j}U_{j+2}U_{j+1}(2\Omega_{q}-U_{j})(2 \Omega_{q} -U_{j+2}) \;=\;0.
\end{equation}

\noindent
The generators $\check{R}_{j} = \frac{q-q^{-1}}{2} + H_{j}(q,1)$ satisfy the
braiding relations

\begin{equation}
\label{e14}
\check{R}_{j} \check{R}_{j\pm 1} \check{R}_{j} \;=\;
\check{R}_{j\pm 1} \check{R}_{j} \check{R}_{j\pm 1}
\end{equation}

\noindent
with

\begin{equation}
\label{e15}
\check{R}_{j}^{2} \;=\; (q-q^{-1})\:\check{R}_{j}\:+\: 1.
\end{equation}

\noindent
Considering the matrices $R_{j}=P \check{R}_{j}$ ($P$ is the graded
permutation operator) we have \\
(see~Eq.~(\ref{e11}))

\begin{equation}
\label{e16}
R \Delta(\alpha)R^{-1} \;=\; \Delta(\alpha^{-1}).
\end{equation}

\noindent
We now consider the case $\eta \neq 1$. As suggested by Eqs.
(\ref{v9},\ref{v10}) we define the two parameter deformation of the
$SU(1/1)$ algebra as follows:

\begin{eqnarray}
\label{e17}
 & \{T^{X},T^{X}\} \;=\; 2\:[E]_{\alpha}, \ \ \ \ \ \ \ \
 & \{T^{Y},T^{Y}\} \;=\; 2\:[E]_{\beta} \\
 & \{T^{X},T^{Y}\} \;=\; 0 \ \ \ \ \ \ \ \ \ \ \ \ \ \ \
 & [E,T^{X}] \;=\; [E,T^{Y}] \;=\; 0 \nonumber
\end{eqnarray}

\noindent
with the coproduct

\begin{eqnarray}
\label{e18}
& & \Delta(\alpha,\beta;T^{X}) \;=\; \alpha^{E/2}\otimes T^{X} \;+\;
T^{X}\otimes\alpha^{-E/2} \nonumber \\
& & \Delta(\alpha,\beta;T^{Y}) \;=\; \beta^{E/2}  \otimes T^{Y} \;+\;
T^{Y}\otimes\beta^{-E/2} \\
& & \Delta(\alpha,\beta;E) \;=\; E\otimes {\bf1} \;+\; {\bf1} \otimes E.
\nonumber
\end{eqnarray}

Notice that $S^{z}$ does not appear in the algebra anymore. We denote this
quantum algebra by $U_{\alpha,\beta}[SU(1/1)]$. It is a Hopf algebra for the
sam
reasons as the $U_{\alpha}[SU(1/1)]$. If we take the fermionic
representations $E=1$, \ $\tau^{x}=(a^{+}+a^{-})$, \
\ $\tau^{y}=-i(a^{+}-a^{-})$, from Eq. (\ref{e18}) we derive
Eqs. (\ref{v9},\ref{v10}). The
quantum chain $H(q,\eta)$ is  thus invariant under the quantum algebra
$U_{\alpha,\beta}[SU(1/1)]$. We would like to see what replaces the relations
(\ref{e12}-\ref{e16}) when we have two parameters. We first notice a remarkable
identity satisfied by the $H_{j}(q,\eta)$

\begin{eqnarray}
\label{e19}
& & \left[ H_{j}H_{j\pm1}H_{j}-H_{j\pm1}H_{j}H_{j\pm1} + (\nu-1)
(H_{j}-H_{j\pm1}) \right] \:(H_{j}-H_{j\pm1}) = \mu \\ \nonumber \\
& & \ \ \ \ \ \ \ \ \ \ \ \ \ \ \ \ \ \ \ \  \ \
\ \ \ \ \ \ \ \ \ \ \ \ \ H_{j}^{2} \;=\; \nu, \nonumber
\end{eqnarray}

\noindent
where

\begin{eqnarray}
\label{e20}
& & \nu \;=\; \left(\frac{\alpha+\alpha^{-1}}{2}\right)
    \left(\frac {\beta+\beta^{-1}}{2}\right)
\;=\; {\left(\frac{q+q^{-1}}{2}\right)}^{2} +
    {\left(\frac{\eta+\eta^{-1}}{2}\right)}^{2} -1 \\
& & \mu \;=\;
    {\left( \frac{\alpha+\alpha^{-1}}{2}-
       \frac{\beta+\beta^{-1}}{2} \right)}^{2} \;=\;
    4\:{\left(\frac{q-q^{-1}}{2}\right)}^{2}\;
    {\left(\frac{\eta-\eta^{-1}}{2}\right)}^{2}.
\nonumber  \end{eqnarray}

\noindent
We can now define a generalised Hecke algebra taking

\begin{eqnarray}
\label{e21}
& & U_{i} \;=\; \sqrt{\nu}-H_{i}(p,q) \nonumber \\
& & (U_{i}U_{i\pm1}U_{i}-U_{i\pm1}U_{i}U_{i\pm1}-U_{i}+U_{i\pm1}) \;
(U_{i}-U_{i\pm1}) \;=\; \mu \\
& & U_{i}^{2} \;=\; 2\:\sqrt{\nu}\:U_{i}. \nonumber
\end{eqnarray}

Notice that when $\eta=1$, \ $\mu=0$ and we are back to the original Hecke
algeb
We did not have the patience to find the equivalent of Eq. (\ref{e13}) which
gives the quotient of the generalised Hecke algebra (\ref{e21}) corresponding
to
the chain given by Eq. (\ref{e4}). Another quotient is however suggested by the
structure of Eq. (\ref{e21}):

\begin{equation}
\label{e22}
(U_{i}U_{i\pm1}U_{i}-U_{i})\:(U_{i}-U_{i\pm1}) \;=\; \frac{\mu}{2}.
\end{equation}

For $\mu=0$ one gets in this case the Temperley-Lieb algebra. We now turn our
attention to the generalised braid group algebra. We take
$\check{R}_{i} = H_{i}(q,\eta)+\sqrt{\nu-1}$ and get

\begin{equation}
\label{e23}
(\check{R}_{i}\check{R}_{i\pm1}\check{R}_{i} -
 \check{R}_{i\pm1}\check{R}_{i}\check{R}_{i\pm1}) \:
(\check{R}_{i}-\check{R}_{i\pm1}) \;=\; \mu
\end{equation}

\noindent
with

\begin{equation}
\label{e24}
\check{R}_{i}^{2} \;=\; 1+\sqrt{\nu-1}\:\check{R}_{i}\;.
\end{equation}

\noindent
In the basis where the $\sigma_{i}^{z}$ are diagonal (see Eq. (\ref{e4}))
we have

\begin{equation}
\label{e25}
\check{R}_{i} \;=\; \left(
\begin{array}{cccc}
\sqrt{\nu-1}+\frac{q+q^{-1}}{2}  &  0  &  0  &  \frac{\eta-\eta^{-1}}{2} \\
0 & \sqrt{\nu-1}-\frac{q-q^{-1}}{2} & \frac{\eta+\eta^{-1}}{2} & 0 \\
0 & \frac{\eta+\eta^{-1}}{2} & \sqrt{\nu-1}+\frac{q+q^{-1}}{2} & 0 \\
\frac{\eta-\eta^{-1}}{2} & 0 & 0 & \sqrt{\nu-1}-\frac{q+q^{-1}}{2}
\end{array}
\right).
\end{equation}

\noindent
We take the graded permutation matrix P

\begin{equation}
\label{e26}
P \;=\; \left(
\begin{array}{cccc}
\;1 & & & \\
& \;0 & \;1 & \\
& \;1 & \;0 & \\
& & & -1
\end{array}
\right)
\end{equation}

\noindent
and define the matrix $R_{i}=P\check{R}_{i}$. We now write the coproduct
(\ref{e18}) in the original language of Pauli matrices

\begin{eqnarray}
\label{e27}
& & \Delta(\alpha,\beta;T^{X}) \;=\;
\alpha^{-1/2}(\sigma^{y}\otimes\bf{1}) \;-\;
\rm{\alpha^{1/2} (\sigma^{z}\otimes\sigma^{y})} \nonumber \\
& & \Delta(\alpha,\beta;T^{Y}) \;=\;
\beta^{-1/2}(\sigma^{x}\otimes\bf{1}) \;-\;
\rm{\beta^{1/2} (\sigma^{z}\otimes\sigma^{x})} \\
& & \Delta(\alpha,\beta;E) \;=\; \bf{1}\otimes\bf{1} \;+\; \bf{1}\otimes\bf{1}.
\nonumber
\end{eqnarray}

\noindent
It is trivial to check that similar to Eq. (\ref{e16}) we get

\begin{equation}
\label{e28}
R\:\Delta(\alpha,\beta)\:R^{-1} \;=\; \Delta(\alpha^{-1},\beta^{-1}).
\end{equation}

Before concluding we would like to show that for $U_{\alpha,\beta}[SU(1/1)]$
one can introduce more than two parameters (as in the Lie algebra case when we
had more than one). The most general chain which has a zero mode for all its
values of the parameters is \cite{r5}

\begin{eqnarray}
\label{e29}
H_{i}\;\;=\; & \;\frac{1}{2} \: \{ \: &
\frac{\Theta+\Theta^{-1}}{2} (\eta\zeta\sigma_{i}^{x}\sigma_{i+1}^{x}
+ \eta^{-1}\zeta^{-1}\sigma_{i}^{y}\sigma_{i+1}^{y}) \nonumber \\
& \;\;+ & \frac{\Theta-\Theta^{-1}}{2}
(\eta\zeta^{-1}\sigma_{i}^{x}\sigma_{i+1}^{y} +
\eta^{-1}\zeta\sigma_{i}^{y}\sigma_{i+1}^{x}) \\
& \;\;+ & q\:\sigma_{i}^{z} \;+\;
q^{-1}\sigma_{i+1}^{z}) \: \}. \nonumber
\end{eqnarray}

\noindent
$H_{i}$ depends on four parameters. One can check however that Eq. (\ref{e19})
holds with

\begin{eqnarray}
\label{e30}
\nu \; & = & \; \frac{q^{2}+q^{-2}}{4} +
				\frac{(\eta^{2}-\eta^{-2})(\zeta^{2}-\zeta^{-2})}{8} +
\frac{(\eta^{2}+\eta^{-2})(\zeta^{2}+\zeta^{-2})(\Theta^{2}+\Theta^{-2})}{16}
    \nonumber \\
    \nonumber \\
\mu \; & = & \;
    \left(
    \frac{(\eta^{2}-\eta^{-2})(\zeta^{2}-\zeta^{-2})}{4} +
    \frac{(\eta^{2}+\eta^{-2})(\zeta^{2}+\zeta^{-2})
    (\Theta^{2}+\Theta^{-2})}{8} - 1
    \right)
    \\
& & \ \times \ \
    \left(
    \frac{{(q-q^{-1})}^{2}}{2} +
    \frac{{(\Theta-\Theta^{-1})}^{2}
    {(\frac{\eta}{\zeta}-\frac{\zeta}{\eta})}^{2}}{8}
    \right)
    \nonumber
\end{eqnarray}

\noindent
which means that we are back to two parameters.
\vspace{5mm} \\
To sum up we have shown that the two-parameter deformation of the $SU(1/1)$
superalgebra leads to a new Hopf algebra. One obtains then a generalisation
of the Hecke algebra and of the braid group (see Eqs. (\ref{e21}) and
(\ref{e23}
keeping the structure of connection between the R~matrices and the coproduct
(see Eq. (\ref{e28})).
\vspace{10mm} \\
Acknowledgements: We would like to thank Franz Gaehlen for writing the computer
program which made us discover Eq. (\ref{e19}). We would also like to thank D.
Altschuler, L. Alvarez-Gaum\'{e}, D. Arnaudon, M. Chaichian, R. Coqueraux,
C. Cinkovic and M. Scheunert for usefull discussions.

\end{document}